# TEXT STEGANOGRAPHIC APPROACHES: A COMPARISON

Monika Agarwal

Department of Computer Science and Engineering, PDPM-IIITDM, Jabalpur, India
peace1287@gmail.com

## ABSTRACT

*This paper presents three novel approaches of text steganography. The first approach uses the theme of missing letter puzzle where each character of message is hidden by missing one or more letters in a word of cover. The average Jaro score was found to be 0.95 indicating closer similarity between cover and stego file. The second approach hides a message in a wordlist where ASCII value of embedded character determines length and starting letter of a word. The third approach conceals a message, without degrading cover, by using start and end letter of words of the cover. For enhancing the security of secret message, the message is scrambled using one-time pad scheme before being concealed and cipher text is then concealed in cover. We also present an empirical comparison of the proposed approaches with some of the popular text steganographic approaches and show that our approaches outperform the existing approaches.*

## 1. INTRODUCTION

Steganography is derived from a finding by Johannes Trithemus (1462-1516) entitled "Steganographia" and comes from the Greek words (στεγανό-ς, γραφ-ειν) meaning "covered writing" [1, 2, 3]. Steganography is the art and science of hiding a message inside another message without drawing any suspicion to others so that the message can only be detected by its intended recipient [4]. Cryptography and Steganography are ways of secure data transfer over the Internet. Cryptography scrambles a message to conceal its contents; steganography conceals the existence of a message. It is not enough to simply encipher the traffic, as criminals detect, and react to, the presence of encrypted communications [5]. But when information hiding is used, even if an eavesdropper snoops the transmitted object, he cannot surmise the communication since it is carried out in a concealed way. Limitation of cryptography is that the third party is always aware of the communication because of the unintelligible nature of the text. Steganography overcomes this limitation by hiding message in an innocent looking object called cover. Steganography gained importance because the US and the British government, after the advent of 9/11, banned the use of cryptography and publishing sector wanted to hide copyright marks [6]. Modern steganography is generally understood to deal with electronic media rather than physical objects and texts [7]. In steganography, the text to be concealed is called embedded data. An innocuous medium, such as text, image, audio, or video file; which is used to hide embedded data is called cover. The key (optional) used in embedding process is called stego-key. A stego-key is used to control the hiding process so as to restrict detection and/or recovery of embedded data to the parties who know it [1]. The stego object is an object we get after hiding the embedded data in a cover medium.

This paper presents three novel approaches of text steganography. The first approach uses the theme of missing letter puzzle and hides each character of secret message in a word by missing one or two letters in that word depending on the ASCII value of the embedded character. The second approach works by hiding message in a list of words where starting letter of word and word length is determined by the ASCII value of the character to be hidden. In the third approach, the cover comprising of paragraphs can be drawn from any source like newspaper/book. The approach conceals secret bits using start and end letter of words of cover file. Unlike the first two proposed approaches in which cover comprising of collection of words is dynamically generated, the third approach makes use of pre-existing any meaningful piece of English text as a cover file to hide the secret bits. The message is scrambled by the proposed encipher algorithm using a one-time secret key. The resulting cipher text is then hidden in cover file by an embedding algorithm using a stego key.

The rest of the paper is organized as follows: Section 2 throws light on text steganography and lists some of the popular approaches of text steganography. Section 3 describes the proposed approaches. Section 4 shows the results of comparison of the proposed methods with the existing methods. Section 5 discusses the merits and demerits of the proposed approaches and other related issues. Section 6 draws the conclusion.

## 2. TEXT STEGANOGRAPHY

Steganography can be classified into image, text, audio and video steganography depending on the cover media used to embed secret data. Text steganography can involve anything from changing the formatting of an existing text, to changing words within a text, to generating random character sequences or using context-free grammars to generate readable texts [7]. Text steganography is believed to be the trickiest due to deficiency of redundant information which is present in image, audio or a video file. The structure of text documents is identical with what we observe, while in other types of documents such as in picture, the structure of document is different from what we observe. Therefore, in such documents, we can hide information by introducing changes in the structure of the document without making a notable change in the concerned output [8]. Unperceivable changes can be made to an image or an audio file, but, in text files, even an additional letter or punctuation can be marked by a casual reader [9]. Storing text file require less memory and its faster as well as easier communication makes it preferable to other types of steganographic methods [10]. Text steganography can be broadly classified into three types: Format based Random and Statistical generation, Linguistic methods.

### Format Based Methods

Format based methods involve altering physically the format of text to conceal the information. This method has certain flaws. If the stego file is opened with a word processor, misspellings and extra white spaces will get detected. Changed fonts sizes can arouse suspicion to a human reader. Additionally, if the original plaintext is available, comparing this plaintext with the suspected steganographic text would make manipulated parts of the text quite visible [7].

### Random and Statistical Generation

In order to avoid comparison with a known plaintext, steganographers often resort to generating their own cover texts [7]. One method is concealing information in random looking sequence of characters. In another method, the statistical properties of word length and letter frequencies are used in order to create words which will appear to have same statistical properties as actual words in the given language [2, 3].

### Linguistic Steganography

Linguistic steganography specifically considers the linguistic properties of generated and modified text, and in many cases, uses linguistic structure as the space in which messages are hidden [7]. CFG create tree structure which can be used for concealing the bits where left branch represents '0' and right branch corresponds to '1'. A grammar in GNF can also be used where the first choice in a production represents bit 0 and the second choice represents bit 1. This method has some drawbacks. First, a small grammar will lead to lot of text repetition. Secondly, although the text is syntactically flawless, but there is a lack of semantic structure. The result is a string of sentences which have no relation to one another [7].

## 2.1. Existing Approaches

In this sub-section, we present some of the popular approaches of text steganography.

### 2.1.1. Line Shift

In this method, secret message is hidden by vertically shifting the text lines to some degree [10, 11]. A line marked has two unmarked control lines one on either side of it for detecting the direction of movement of the marked line [12]. To hide bit 0, a line is shifted up and to hide bit 1, the line is shifted down [13]. Determination of whether the line has been shifted up or down is done by measuring the distance of the centroid of marked line and its control lines [12]. If the text is retyped or if a character recognition program (OCR) is used, the hidden information would get destroyed. Also, the distances can be observed by using special instruments of distance assessment [10].

### 2.1.2. Word Shift

In this method, secret message is hidden by shifting the words horizontally, i.e. left or right to represent bit 0 or 1 respectively [13]. Words shift are detected using correlation method that treats a profile as a waveform and decides whether it originated from a waveform whose middle block has been shifted left or right [12]. This method can be identified less, because change of distance between words to fill a line is quite common [10, 11]. But if someone knows the algorithm of distances, he can compare the stego text with the algorithm and obtain the hidden content by using the difference. Also, retyping or using OCR programs destroys the hidden information [10, 11].

### 2.1.3. Syntactic Method

This technique uses punctuation marks such as full stop (.), comma (,), etc. to hide bits 0 and 1. But problem with this method is that it requires identification of correct places to insert punctuation marks [10, 11]. Therefore, care should be taken in using this method as readers can notice improper use of the punctuations [9].

### 2.1.4 White Steg

This technique uses white spaces for hiding a secret message. There are three methods of hiding data using white spaces. In Inter Sentence Spacing, we place single space to hide bit 0 and two spaces to hide bit 1 at the end of each terminating character [9]. In End of Line Spaces, fixed number of spaces is inserted at the end of each line. For example, two spaces to encode one bit per line, four spaces to encode two bits and so on. In Inter Word Spacing technique, one space after a word represents bit 0 and two spaces after a word represents bit 1. But, inconsistent use of white space is not transparent [9].

### 2.1.5. Spam Text

HTML and XML files can also be used to hide bits. If there are different starting and closing tags, bit 0 is interpreted and if single tag is used for starting and closing it, then bit 1 is interpreted [13]. In another technique, bit 0 is represented by a lack of space in a tag and bit 1 is represented by placing a space inside a tag [13].

### 2.1.6. SMS-Texting

SMS-Texting language is a combination of abbreviated words used in SMS [8]. We can hide binary data by using full form of word or its abbreviated form. A codebook is made which contains words and their corresponding abbreviated forms. To hide bit 0, full form of the word is used and to hide bit 1, abbreviated form of word is used [8].

### 2.1.7. Feature Coding

In feature coding, secret message is hidden by altering one or more features of the text. A parser examines a document and picks out all the features that it can use to hide the information [13]. For example, point in letters i and j can be displaced, length of strike in letters f and t can be changed, or by extending or shortening height of letters b, d, h, etc. [6, 14]. A flaw of this method is that if an OCR program is used or if re-typing is done, the hidden content would get destroyed.

### 2.1.8. SSCE (Secret Steganographic Code for Embedding)

This technique first encrypts a message using SSCE table and then embeds the cipher text in a cover file by inserting articles a or an with the non specific nouns in English language using a certain mapping technique [15]. The embedding positions are encrypted using the same SSCE table and saved in another file which is transmitted to the receiver securely along with the stego file.

### 2.1.9. Word Mapping

This technique encrypts a secret message using genetic operator crossover and then embeds the resulting cipher text, taking two bits at a time, in a cover file by inserting blank spaces between words of even or odd length using a certain mapping technique [16]. The embedding positions are saved in another file and transmitted to the receiver along with the stego object.

### 2.1.10. MS Word Document

In this technique, text segments in a document are degenerated, mimicking to be the work of an author with inferior writing skills, with secret message being embedded in the choice of degenerations which are then revised with changes being tracked [17]. Data embedding is disguised such that the stego document appears to be the product of collaborative writing [17].

### 2.1.11. Cricket Match Scorecard

In this method, data is hidden in a cricket match scorecard by pre-appending a meaningless zero before a number to represent bit 1 and leaving the number as it is to represent bit 0 [18].

### 2.1.12. CSS (Cascading Style Sheet)

This technique encrypts a message using RSA public key cryptosystem and cipher text is then embedded in a Cascading Style Sheet (CSS) by using End of Line on each CSS style properties, exactly after a semicolon. A space after a semicolon embeds bit 0 and a tab after a semicolon embeds bit 1 [19].

## 3. THE PROPOSED APPROACHES

Figure 1 shows the proposed model of text steganography. The model consists of four building blocks: Encipher function, which enciphers a message using one-time pad scheme, Hide function, which conceals the scrambled message using a stego key, Seek function, which extracts the hidden information from the stego file using the stego key, and, Decipher function, which deciphers the extracted message using the secret key. All the three proposed approaches work on the proposed model and have same encipher and decipher algorithms with different hide and seek algorithms for each approach. In all algorithms, we have assumed integer division, i.e., division by a whole number ignores the fractional part of the result.

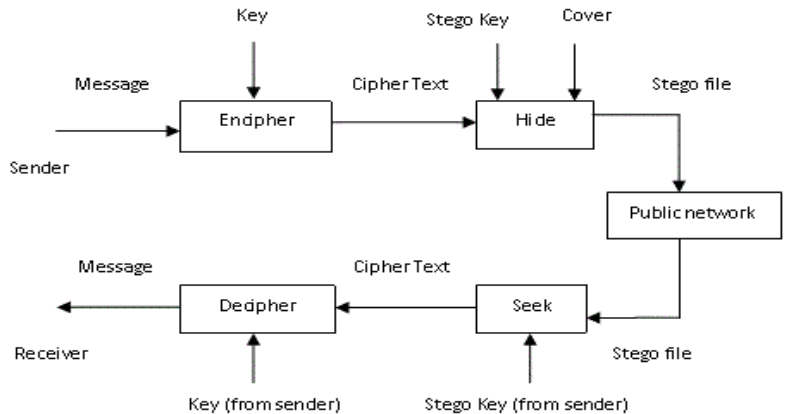

Figure 1. The Proposed Model of Text Steganography

**One-Time Pad**

In this scheme, the key, comprising of true random numbers, used for enciphering a message is as long as the message. In addition, a key is used for single encryption-decryption. Thus, there is no fixed pattern which an eavesdropper can analyse. Such a scheme is unbreakable as it produces output that bears no statistical relationship to the plain text. But one-time pad is of limited utility and is useful for applications requiring very high security [20]. In steganography, security of message is of utmost importance. Therefore, for enhancing the security of the secret message, the concept of one-time pad has been used in the proposed model of text steganography.

**The Encipher Algorithm**

1. Fill an array, A, of size 1000 with random numbers in the range 0 to 255.
2. Read a character of the secret message and get its decimal equivalent (n).
3. Generate a random index, i, to the array A and get a random number, r = A[i].
4. Write r in the key file and generate another random value in the same range and replace it with A[i].
5. s = sum of square of the digits of r.
6. x = s/10,  y = s (mod10)
7. The scrambled value, e = n - (x*y) + r.
8. Write the character equivalent of e in the cipher text file.
9. Repeat steps 2 to 8 till the end of the message.
10. Send the key to the receiver.

**The Decipher Algorithm**

1. Read a character from the input file and get its decimal equivalent, e.
2. Read a value (r) from the key file.
3. s = sum of square of the digits of r.
4. x = s/10, y = s (mod10)
5. Deciphered value, n = e – r + (x*y).
6. Convert n to its character equivalent.
7. Repeat steps till the end of the file.

### 3.1. The First Approach: Missing Letter Puzzle

Missing letter puzzle is a puzzle comprising of a collection of words with one or more letters missing in each word. A letter, at some position in a word, is missed by replacing it with a question mark. The puzzle is solved by replacing each question mark by an appropriate letter in each word so as to make the words meaningful. Words in a puzzle can be of different length and different domains, i.e., they can be terminologies of various fields or can be proper nouns or a combination of both.

In the following proposed approach, we have incorporated the theme of missing letter puzzle. Each character of secret message is hidden in a word of certain length by missing one or two letters in it. Hint is also given with some words. Since the length of words depends on the decimal (ASCII) value of characters to be hidden, the words are dynamically generated and there is no pre-determined cover file.

**Hide Algorithm**

1. Calculate length (ln) of the input file.
2. Read a character from the file and get its decimal equivalent (n).
3. If n<100,
   a) Flag=0
   b) q = n/10 , r = n (mod 10)
   c) If q < 6 then l = 10 + q, else l = q.
   d) Read a l-letter word.
   e) If r = 0 then miss the character at any random position (< = l) of the word and give a hint with that word.
   f) Else if r < = q then miss the character at $r^{th}$ position of the word.
   g) Else miss the character at position (r-q) of the word and also at any random position, p ( if q ϵ {6,7,8,9} then 3 < p <= l; if q ϵ {0,1,…,5} then 9 < p <= l).
4. Else
   a) flag = 1 + least significant digit of n.
   b) q = most significant digit of n
   c) r = middle digit of n
   d) l = 10 + q
   e) Read a l-letter word.
   f) If r = 0, miss the character at random position, p (10 <= p <= l) of the word.
   g) Else miss the character at position r of the word.
5. Write the word in the stego file and the flag in the stego key file.
6. Repeat steps 2 to 5 till the end of the file.
7. If ln < 10 then insert 10 – ln ten letter words (missing the character at random position of the word) in the stego file.
8. The stego file and stego key are sent, separately, to the receiver.

**Seek Algorithm**

1. Read a value (k) from the stego key.
2. Read a word from the stego file.
3. If k=0
   a) Calculate length (l) of the word.
   b) If l > 9, then l = l – 10.
   c) If in the stego file, there is a hint with this word, then r = 0.
   d) Else if there are two missing characters in the word, then r = l + position of first missing character in the word.
   e) Else r = position of the missing character in the word.

  f)   The extracted value, asc = (l*10) + r.
4.  Else
    a)   a = k−1.
    b)   Calculate length (l) of the word.
    c)   l = l−10.
    d)   r = position of the missing character in the word.
    e)   If r > 9 then r = 0.
    f)   The extracted value, asc = (l * 100) + (r * 10) + a.
5.  Convert asc to its character equivalent.
6.  Repeat above steps till the end of the stego key.

**Example**

Consider a message "Hello World!" to be hidden in a cover file. The cipher text of the message was found to be "ĕxŝĩv!ǨÀ¶ĊtP". Since a word is dynamically generated and then a character is hidden in it using a question mark/ hint, in order to measure the amount of degradation, we have assumed that cover consists of set of words without question marks and hints. Figure 2 shows the generated cover file (C), Figure 3 shows the stego file (S).

```
PENTAPRAZOLE
TETRACYCLINE
LEISHMANIASIS
NONIMIDAZOLE
ZAFIRLUKAST
NORTRIPTYLINE
ZOPIDEM
VENLAFAXINE
TICLOPIDINE
CARISOPRODOL
PHENINDIONE
PROPOFOL
```

Figure 2. The Cover File (C)

```
PENTAPR?ZOLE
TETRACYCL?NE    (antibiotic)
LEI?HMANIASIS
NO?IMIDAZOLE
ZAFIRLU?AST
NO?TRIPTYLINE
ZOPI?EM
VENLAFAX?NE
TICLOPI?INE
CARIS?PRODOL
PHENI?DIONE
PROP?FOL    (drug)
```

Figure 3. The Stego File (S)

## 3.2. The Second Approach: Hiding Data in Wordlist

A drawback of the previous approach is that missing letter puzzle is not so common and stego file consists of too many question marks. This proposed method addresses this issue and conceals a message in a list of words without using any special character. Each character is hidden in a word of certain length. The starting letter of word is determined by masking sum of the digits of ASCII value of the character to an English alphabet. If sum of the digits is 1, then starting letter of word will be 'a'; if it is 2, then 'b' and so forth. Since the length and staring letter of words depends on the decimal value of the embedded characters, cover is dynamically generated.

**Hide Algorithm**

1. Calculate length (le) of the input file.
2. Read a character from the file and get its ASCII value (n).
3. If n<100, then make it a three digit number by pre-appending it with zeroes.
4. k = most significant digit of n. Write k in the stego key file.
5. l = middle digit of n.
6. If l < 6, then l = l+10.
7. s = sum of the digits of n.
8. Get a l-letter word starting with the s[th] letter of the English alphabet and write it in the stego file.
9. Repeat steps 2 to 8 till the end of the file.
10. If le < 10, insert 10-le ten letter words in the stego file.

**Seek Algorithm**

1. Read a value (k) from the stego key file.
2. Read a word from the stego file and get its length (l).
3. If l > 9, then l = l-10.
4. Calculate s by decoding the first letter of the word from the English alphabet.
5. r = s - (l + k).
6. The extracted value, n = (k * 100) + (l * 10) + r.
7. Convert n to its character equivalent.
8. Repeat above steps till the end of the stego key file.

**Example**

Consider a message "pascal" to be hidden in a cover file. The cipher text of the message was found to be "ÎdŘÚĞ▯". Figure 4 shows the stego file after hiding the cipher text.

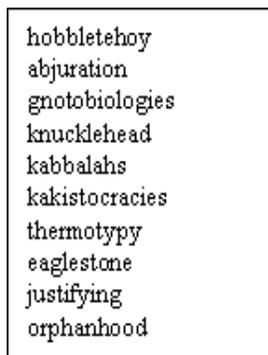

Figure 4. The Stego File

## 3.3. The Third Approach: Hiding Data in Paragraphs

Unlike the two aforementioned proposed approaches in which cover is dynamically generated and stego file consists of a list of words, this approach makes use of a pre-determined cover file which can be any meaningful piece of English text and can be drawn from any source (For example, a paragraph from a newspaper/book). The approach works by hiding a message using start and end letter of the words of a cover file. This approach works on the binary value of a character as opposed to the above two proposed approaches which work on the ASCII value. After converting the cipher text to a stream of bits, each bit is hidden by picking a word from the cover file and using either the start or the end letter of that word depending on the bit to be concealed. Bit 0 or 1 is hidden by reading a word, sequentially, from the cover file and including the starting letter or the end letter, respectively, of the word in the stego key. A word having same start and end letter is skipped. Since no change is made to the cover, the cover file and its corresponding stego file are exactly the same.

**Hide Algorithm**

1. Get a cover file.
2. Convert the input file to its binary equivalent (bin).
3. Read a bit (x) from the bin.
4. Read a word from the cover file and write it in stego file.
5. If start and end letter of the word is same, then read the next word of the cover file and write it in the stego file.
6. s = start letter of the word and e = end letter of the word.
7. If x = 0, write s in the stego key.
8. Else if x = 1, write e in the stego key.
9. Repeat steps 3 to 8 till the end of the bin file.
10. Send the stego file and the stego key to the receiver.

**Seek Algorithm**

1. Read a character (c) from the stego key.
2. Read a word from the stego file.
3. If start and end letter of the word is same, then skip that word and read the next word from the stego file.
4. Get the start letter (s) and end letter (e) of the word.
5. If c = s, then bit b = 0.
6. Else if c = e, then bit b = 1.
7. Write b in a file.
8. Execute above steps repeatedly till the end of the stego key.
9. Convert the file into its character equivalent.

**Example**

Consider a message "try" to be concealed in a cover file. After enciphering the message, cipher text generated was "ωμĉ". Figure 5 shows the cover file and Figure 6 shows the stego file. We can see that the two files are exactly the same.

> The house itself looked empty. The doors and windows were locked. The front verandah bare. Unfurnished. But the skyblue Plymouth with chrome tailfins was still parked outside, and inside, Baby Kochamma was still alive. She was Rahel's baby grandaunt, her grandfather's younger sister. Her name was really Navomi, Navomi Ipe, but everybody called her Baby. She became Baby Kochamma when she was old enough to be an aunt.

Figure 5. The Cover File

> The house itself looked empty. The doors and windows were locked. The front verandah bare. Unfurnished. But the skyblue Plymouth with chrome tailfins was still parked outside, and inside, Baby Kochamma was still alive. She was Rahel's baby grandaunt, her grandfather's younger sister. Her name was really Navomi, Navomi Ipe, but everybody called her Baby. She became Baby Kochamma when she was old enough to be an aunt.

Figure 6. The Stego File

## 4. EXPERIMENTAL RESULTS

This section shows the results of the experiments conducted to measure the performance of the proposed approaches and compares the proposed approaches with the existing text steganographic approaches.

### Capacity

Capacity is defined as the ability of a cover media to hide secret information. The capacity ratio is computed by dividing the amount of hidden bytes over the size of the cover text in bytes [21].

**Capacity ratio = (amount of hidden bytes) / (size of the cover text in bytes)**

Assuming one character occupies one byte in memory, we have calculated the percentage capacity which is capacity ratio multiplied by 100.

The samples of embedded data used are:

1. Ego (3 byte)

2. Minute (6 byte)

3. Hello World! (12 byte)

4. Failure is never final ! (24 byte)

5. Smile is an inexpensive way to improve your looks. (50 byte)

6. Its not the load that breaks you down, its the way you carry it. (63 byte)

7. Don't find hundred reasons why you can't do a thing, but just find one reason why you can and do it. (100 byte)

8. Tide recedes and leaves behind bright sea shells on sand
   Sun sets but its warmth lingers on land
   Music stops and its echoes on in sweet refrains
   For every joy that passes, something beautiful remains (202 byte)

9. Steganography is not a new area. It dates back to $5^{th}$ century BC. Harpagus used hare to send his message by killing it and hiding the message inside its belly. A person disguised as hunter carried the hare to the destination. Another incident was of King Darius of Susa. Histiaeus was assigned the duty of shaving the head of his most trusted slave. (349 byte)

10. Steganography is not a new area. It dates back to $5^{th}$ century BC. Harpagus used hare to send his message by killing it and hiding the message inside its belly. A person disguised as hunter carried the hare to the destination. Another incident was of King Darius of Susa. Histiaeus, prisoner of Darius, was assigned the duty of shaving the head of his most trusted slave and then the message was tattooed on his shaved scalp. After some time, when the hairs of the slave grew back, his head was shaved again. (508 byte)

Table 1 shows the observed percentage capacity of the proposed approaches over the above ten experimental samples. Table 2 compares the average percentage capacity of the proposed approaches with those existing text steganographic approaches for which the observed average percentage capacity is found to be greater than one.

|  | I | II | III | IV | V | VI | VII | VIII | IX | X |
|---|---|---|---|---|---|---|---|---|---|---|
| Missing Letter App. | 2.88 | 5.40 | 9.09 | 8.79 | 8.60 | 8.71 | 8.83 | 8.63 | 8.74 | 8.66 |
| Wordlist Approach | 2.94 | 5.55 | 8.33 | 8.79 | 8.84 | 8.95 | 8.93 | 9.00 | 8.73 | 8.92 |
| Paragraph Approach | 1.96 | 1.92 | 2.06 | 2.00 | 2.08 | 2.12 | 2.13 | 2.17 | 2.16 | 2.15 |

Table 1. Percentage capacity of the proposed approaches over the ten experimental samples.

Table 2. Average percentage capacity of the approaches (having av. % capacity > one).

| Missing Letter App. | Wordlist App. | Paragraph App. | White Steg | SMS Texting | Feature Coding | Word Map | Spam Text | Word Shift |
|---|---|---|---|---|---|---|---|---|
| 7.833 | 7.898 | 2.075 | 1.874 | 1.71 | 1.479 | 1.464 | 1.164 | 1.03 |

## Similarity Measure

Similarity metrics are used to measure distance between two values or sequences. There are four similarity metrics [22]:

*1. Needlemen-Wunch*
This algorithm is used in Bioinformatics for aligning globally two protein sequences.

*2. Champan Ordered Name Compound Similarity*

Chapman Ordered Name Compound Similarity tests similarity upon the most similar terms of token-based name where later name are valued higher than earlier names [22].

*3. Smith-Waterman*

Using this algorithm, similar regions are determined locally between two protein sequences.

*4. Jaro-Winkler*

The similarity between two strings can be calculated by using Jaro-Winkler metric.

**Jaro Winkler Similarity Metric**

Among the four metrics mentioned above, the Jaro-Winkler metric is applicable for measuring similarity between the cover file and stego file generated by using the proposed Missing letter puzzle approach because it deals with strings. The Jaro-Winkler metric (or Jaro score), which is a variant of the Jaro distance, is calculated as,

**Jaro-Winkler(s1, s2) = Jaro(s1, s2) + (L*p (1-Jaro(s1,s2))**

$$\text{Jaro}(s1, s2) = \frac{1}{3} * \left( \frac{m}{|s1|} + \frac{m}{|s2|} + \frac{m-t}{m} \right).$$

Where s1 and s2 are the two strings whose similarity is to be measured, L is the length of common prefix (maximum 4 characters), p is scaling factor whose standard value is 0.1, m is the number of matching characters and t is the number of transpositions [16, 22]. Half the number of matching (but different sequence order) characters defines the number of transpositions required. Two characters from s1 and s2 are said to be matching only if they are not farther than ⌊ max (|s1|, |s2|) / 2⌋ -1. Each character of s1 is compared with all its matching characters in s2 [16]. A high Jaro score represents greater similarity between the strings. Jaro score 0 indicates that the strings are not similar and 1 represents that they are exactly same.

Consider a message "Hello World!" to be hidden in a cover file using the Missing letter approach. The cipher text of the message was found to be "ĕxśîv!KÀ¶ĊtP". Table 3 shows the Jaro score of the string pairs when the Jaro-Winkler metric is applied to the cover file (C) and the stego file (S) shown by the Figure 2 and the Figure 3 respectively. The average Jaro score was found to be 0.93.

Table 3. Jaro score of the string pairs of the cover file (C) and the stego file (S).

| String Pair | CS$_1$ | CS$_2$ | CS$_3$ | CS$_4$ | CS$_5$ | CS$_6$ | CS$_7$ | CS$_8$ | CS$_9$ | CS$_{10}$ | CS$_{11}$ | CS$_{12}$ |
|---|---|---|---|---|---|---|---|---|---|---|---|---|
| Jaro Score | 0.96 | 0.85 | 0.96 | 0.95 | 0.96 | 0.95 | 0.94 | 0.96 | 0.96 | 0.96 | 0.96 | 0.84 |

Where CS$_i$ is the ith pair of strings belonging to the cover file (C) and the stego file (S), i.e.,

**CS$_i$ = (C$_i$ , S$_i$)**

C$_i$ is the ith string of the cover file (C) and S$_i$ is the corresponding ith string of the stego file (S).

Table 4 shows the Jaro score of ten cover files and their corresponding stego files generated by using the same approach when the Jaro-Winkler metric is applied to the following ten different samples of embedded data. The average Jaro score of the ten samples of embedded data was found to be 0.95. The samples of embedded data used are:

1. A
2. try
3. smile
4. silence
5. Happiness
6. possibility
7. Steganography
8. glimmer of hope
9. the art of living
10. outstanding success

Table 4. Jaro score of the above ten samples of embedded data.

| Samples | I | II | III | IV | V | VI | VII | VIII | IX | X |
|---|---|---|---|---|---|---|---|---|---|---|
| Jaro Score | 0.952 | 0.961 | 0.956 | 0.952 | 0.953 | 0.960 | 0.962 | 0.956 | 0.958 | 0.958 |

Since in the proposed Wordlist and Paragraph approaches data is hidden without altering the cover, cover and its corresponding stego file are exactly the same leading to the Jaro score of 1.

## 5. DISCUSSION

The three main performance parameters to be considered when studying steganographic systems are: capacity, security and robustness. Capacity refers to the ability of cover media to conceal secret information. Security refers to the ability of an eavesdropper to figure out the hidden information easily. Robustness refers to the ability of protecting unseen data from modification [21].

The observed average percentage capacity of the proposed approaches was found to be better than the existing approaches. The average percentage capacity of the first two proposed approaches is higher than the third approach. The reason is that the former work on ASCII value of embedded character rather than its binary equivalent.

The proposed approaches are secure as they do not make use of extra white spaces or misspelled words to hide secret data. Thus, stego files will not draw suspicion regarding the existence of hidden information if opened with a word processor program. The first two proposed approaches hide each embedded character in a word of cover. But in case of smaller messages, stego files of three or four words will raise suspicion. Therefore, for security, extra words are added in the stego file if secret message is of less than ten bytes and thus, stego file contains at least ten words regardless of the message size. The Wordlist approach is more secure than the Missing Letter approach because it hides data in a set of words without using any special characters. But the stego file generated using the Wordlist approach does not contains any natural looking information and, thus, multiple stego files may draw suspicion. The third proposed approach addresses this issue by hiding data in any natural looking meaningful piece of English Text which can be drawn from any source like newspaper, book, etc. To further enhance security, secret message is first scrambled using one time pad scheme and the resulting cipher text is then hidden in a cover. The proposed encipher algorithm uses a secret key of nearly true random numbers to encrypt a message because pseudo random numbers are not so secure for cryptographic applications and true random numbers cannot be generated by software as they depend on environmental conditions like radioactive decay or CPU temperature, etc. Further, length of key is same as length of message and a key is used for single encryption-decryption. If the same message is encrypted again, a different key is used and thus the cipher text is different. So, there is no definite pattern of the cipher text/stego file which an eavesdropper can analyze to break it. Hence, it is secure.

The proposed approaches are robust as there are no extra white spaces whose removal can destroy the hidden content. Also, the stego files can withstand OCR techniques as no font changing methods have been applied and retyping does not destroy the concealed content.

The first two proposed approaches do not have a pre-determined cover. We have assumed that the set of generated words comprise the cover file. The average Jaro score using the Missing letter approach was found to be 0.95 indicating that cover and its stego file were closely similar. As the Wordlist approach does not alter the words of cover, cover and its stego file are exactly the same leading to the Jaro score of 1. Similarly, as the Paragraph approach hides data using start/end letter of the words of cover file without degrading it, both the files are exactly the same. So, even if an eavesdropper gets both the cover and the stego file, he cannot figure out any difference in the two files. Table 5 compares the three proposed approaches. From the table it is evident that the third approach is most efficient.

Table 5. Comparison of the three proposed approaches.

| Proposed Approaches | Cover | Works on | Av. % Capacity | Security | Av. Jaro Score |
|---|---|---|---|---|---|
| Missing Letter Puzzle | Missing Letter Puzzle | ASCII value | 7.833 | Secure | 0.95 |
| Wordlist Approach | List of Words | ASCII value | 7.898 | More secure | 1 |
| Paragraph Approach | Any meaningful piece of English text | Binary value | 2.075 | Most secure | 1 |

## 6. CONCLUSIONS

A coin has two sides. Steganography although conceals the existence of a message but is not completely secure. It is not meant to supersede cryptography but to supplement it. This paper presents three novel approaches of text steganography. The first approach uses the theme of missing letter puzzle; the second hides a message in a list of related words, and the third approach works by using start and end letter of words of any natural looking meaningful piece of English text. Embedding is done in such a way that there is no or minimum degradation of the cover. The average Jaro score for the Missing Letter approach was found to be 0.95 which indicates closer similarity between the cover and stego files. As the second and third proposed approaches conceal a message without altering cover, their Jaro score is one. The observed average percentage capacity of the proposed approaches was found to be better than the existing approaches. As the stego files do not contain any extra white space or misspelled words, they do not draw suspicion when opened with a word processor program and retyping does not lead to the loss of hidden content. Thus, the files are secure and robust. The approaches allow further security by scrambling a message, using one-time pad scheme, before concealing it and thereby leading the message security equivalent to the most secure one-time pad system. As no font altering methods have been applied, the stego files can withstand OCR techniques. The last approach being the most efficient, can be used to transmit confidential data such as user passwords, PIN, etc. securely over the Internet. Also, the concept of this approach can also be applied, with little modification, to other languages.

## REFERENCES

[1] F. A. P. Petitcolas, R.J. Anderson, and M. G. Kuhn, "Information hiding- a survey," In *Proceedings of IEEE*, vol.87, pp. 1062-1078, 1999.

[2] L. Y. Por, and B. Delina, "Information hiding- a new approach in text steganography," $7^{th}$ *WSEAS Int. Conf. on Applied Computer and Applied Computational Science*, 2008, pp. 689-695.

[3] L. Y. Por, T. F. Ang, and B. Delina, "WhiteSteg- a new scheme in information hiding using text steganography," *WSEAS Transactions on Computers,* vol.7, no.6, pp. 735-745, 2008.

[4] S. Changder, D. Ghosh, and N. C. Debnath, "Linguistic approach for text steganography through Indian text," 2010 $2^{nd}$ *Int. Conf. on Computer Technology and Development*, 2010, pp. 318-322.

[5] R.J. Anderson, and F. A. P. Petitcolas, "On the limits of steganography," *IEEE Journal of Selected Areas in Communication*, vol.16, pp. 474-481, 1998.

[6] K. Rabah, "Steganography-the art of hiding data," *Information Technology Journal*, vol.3, pp. 245-269, 2004.

[7] K. Benett, "Linguistic steganography- survey, analysis and robustness concerns for hiding information in text," Purdue University, CERIAS Tech. Report 2004-13, 2004.

[8] M. S. Shahreza, and M. H. S. Shahreza, "Text steganography in SMS," 2007 *Int. Conf. on Convergence Information Technology*, 2007, pp. 2260-2265.

[9] W. Bender, D. Gruhl, N. Morimoto, and A. Lu, "Techniques for data hiding," *IBM Systems Journal*, vol.35, pp. 313-336, 1996.

[10] M. H. S. Shahreza, and M. S. Shahreza, "A new approach to Persian/Arabic text steganography," In Proceedings of $5^{th}$ *IEEE/ACIS Int. Conf. on Computer and Information Science and $1^{st}$ IEEE/ACIS Int. Workshop on Component-Based Software Engineering, Software Architecture and Reuse*, 2006, pp. 310-315.

[11] M. H. S. Shahreza, and M. S. Shahreza, "A new synonym text steganography," *Int. Conf. on Intelligent Information Hiding and Multimedia Signal Processing*, 2006, pp. 1524-1526.

[12] S. H. Low, N. F. Maxemchuk, J. T. Brassil, and L. O. Gorman, "Document marking and identification using both line and word shifting," *INFOCOM'95 Proceedings of the Fourteenth Annual Joint Conf. of the IEEE Computer and Communication Societies*, 1995, pp. 853-860.

[13] J. Cummins, P. Diskin, S. Lau, and R. Parlett, "Steganography and digital watermarking," School of Computer Science, 2004, pp.1-24.

[14] J. T. Brassil, S. Low, N. F. Maxemchuk, and L. O. Gorman, "Electronic marking and identification techniques to discourage document copying," *IEEE Journal on Selected Areas in Communication*, vol.1, pp. 1495-1504, 1995.

[15] I. Banerjee, S. Bhattacharyya, and G. Sanyal, "Novel text steganography through special code generation," *Int. Conf. on Systemics, Cybernetics and Informatics*, 2011, pp. 298-303.

[16] S. Bhattacharyya, I. Banerjee, and G. Sanyal, "A novel approach of secure text based steganography model using word mapping method," *Int. Journal of Computer and Information Engineering*, vol.4, pp. 96-103, 2010.

[17] T. Y. Liu, and W. H. Tsai, "A new steganographic method for data hiding in Microsoft word documents by a change tracking technique," *IEEE Transactions on Information Forensics and Security*, vol.2, no.1, pp. 24-30, 2007.

[18] M. Khairullah, "A novel text steganography system in cricket match scorecard," *Int. Journal of Computer Applications,* vol.21, pp. 43-47, 2011.

[19] H. Kabetta, B. Y. Dwiandiyanta, and Suyoto, "Information hiding in CSS: a secure scheme text steganography using public key cryptosystem," *Int. Journal on Cryptography and Information Security*, vol.1, pp. 13-22, 2011.

[20]    William Stallings, *Cryptography and Network Security: Principles and Practice 5/e.*, India, Prentice Hall, 2011.

[21]    F. A. Haidari, A. Gutub, K. A. Kahsah, and J. Hamodi, "Improving security and capacity for Arabic text steganography using "kashida" extensions," 2009 *IEEE/ACS Int. Conf. on Computer Systems and Applications*, 2009, pp. 396-399.

[22]    S. D. Pandya, P. V. Virparia, "Testing various similarity metrics and their permutations with clustering approach in context free data cleaning," *Int. Journal of Computer Science and Security*, vol.3, pp. 344-350, 2009.

**Author**

**Monika Agarwal** received her B.Tech degree in Information Technology from G.L.A. Institute of Technology and Management, Mathura, India, now known as G.L.A. University and M.Tech Degree in Computer Science and Engineering from Indian Institute of Information Technology, Design and Manufacturing, Jabalpur, India. Her areas of interest are Steganography and Digital Watermarking.

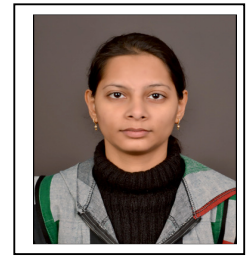